\input epsf.tex
\documentclass[12pt]{article}
\usepackage{graphicx}
\csname @addtoreset\endcsname{equation}{section}

\def\bseq{\begin{subequation}}  
\def\eseq{\end{subequation}}
\def\bsea{\begin{subeqnarray}}  
\def\esea{\end{subeqnarray}}

\newcommand{\bbox}{\lower.2ex\hbox{$\Box$}}

\newcommand{\beq}{\begin{equation}}
\newcommand{\eeq}{\end{equation}}
\newcommand{\bea}{\begin{eqnarray}}
\newcommand{\eea}{\end{eqnarray}}
\newcommand{\ena}{\end{eqnarray}}

\renewcommand{\a}{\alpha}
\renewcommand{\b}{\beta}

\renewcommand{\d}{\delta}

\newcommand{\pa}{\partial}
\newcommand{\g}{\gamma}
\newcommand{\G}{\Gamma}

\newcommand{\D}{\Delta}
\newcommand{\e}{\epsilon}
\newcommand{\z}{\zeta}

\newcommand{\m}{\mu}

\newcommand{\n}{\nu}

\newcommand{\s}{\sigma}
\renewcommand{\S}{\Sigma}

\newcommand{\Db}{\bar{D}}

\newcommand{\Phib}{\bar{\Phi}}

\newcommand{\ad}{{\dot{\alpha}}}

\begin{document}
\begin{titlepage}
\begin{flushright}
IFUM--558/FT \\
KUL--TF--2000/16\\
Bicocca--FT/00/07 \\
\end{flushright}
\vspace{.3cm}
\begin{center}
{\Large \bf More on correlators and contact terms \\
\vskip 5pt
in ${\cal N}=4$ SYM at
order $g^4$ }
\vfill

{\large \bf Silvia Penati$^1$,
Alberto Santambrogio$^2$ and
Daniela Zanon$^3$}\\
\vfill

{\small
$^1$ Dipartimento di Fisica dell'Universit\`a di Milano--Bicocca
and\\ INFN, Sezione di Milano, Via Celoria 16,
20133 Milano, Italy\\
\vspace*{0.4cm}
$^2$  Instituut voor Theoretische Fysica - Katholieke Universiteit Leuven
\\Celestijnenlaan 200D B--3001 Leuven, Belgium\\
\vspace*{0.4cm}
$^3$ Dipartimento di Fisica dell'Universit\`a di Milano
and\\ INFN, Sezione di Milano, Via Celoria 16,
20133 Milano, Italy\\}
\end{center}
\vfill
\begin{center}
{\bf Abstract}
\end{center}
{\small
We compute two-point functions of chiral operators ${\rm Tr} \Phi^k$
for {\em any} $k$, in ${\cal N}=4$ supersymmetric $SU(N)$ Yang-Mills theory.
We find that up to the order $g^4$ the perturbative
corrections to the correlators vanish for {\em all}
$N$. The cancellation occurs in a highly non trivial
way, due to a
complicated interplay between planar and non planar diagrams.\\
In complete generality we  show that this same result is valid for
 {\em any} simple
 gauge group.\\
Contact term contributions signal the presence of ultraviolet
divergences. They are arbitrary at the tree level, but
the absence of perturbative renormalization in the non singular
part of the correlators allows to compute them
unambiguously at higher orders. In the spirit of the AdS/CFT correspondence
we comment on their relation to infrared singularities in the
supergravity sector.}

\vspace{2mm} \vfill \hrule width 3.cm
\begin{flushleft}
e-mail: silvia.penati@mi.infn.it\\
e-mail: alberto.santambrogio@fys.kuleuven.ac.be\\
e-mail: daniela.zanon@mi.infn.it
\end{flushleft}
\end{titlepage}

\section{Introduction}

In this paper we continue the work presented in Ref. \cite{us}. There
we computed two-point functions of chiral operators ${\rm Tr} \Phi^3$
in ${\cal N}=4$ $SU(N)$ supersymmetric Yang-Mills theory to the order
$g^4$ in perturbation theory and proved that corrections vanish
for all values of $N$. In our perturbative approach, the fact that
the non renormalization occurs for any value of $N$, not only at
leading order in the $N\rightarrow \infty$ limit, was strictly
connected to the vanishing of the colour combinatoric factor of the
nonplanar diagrams. The open question was to
see if the same pattern were true also for correlators of operators
${\rm Tr} \Phi^k$, with general $k$.

We have found that for $k>3$ the colour structure of non planar
diagrams at order $g^4$ does not vanish. It is only a complicated
cancellation between $1/N$ subleading contributions from planar
diagrams and nonplanar ones which allows to obtain a
complete {\em all} $N$ nonrenormalization of the two-point
correlators.

The colour structure identities we have used in
our calculations can be generalized to arbitrary gauge groups. In so
doing we have been able to prove that the nonrenormalization of the
correlators up to order $g^4$ is valid for {\em any} group.

The other issue we focus on in this paper, is related to the presence
of contact term contributions \cite{cont}.
They arise in the computation of the
correlators because the theory is affected by ultraviolet
divergences. In order to control such infinities one has to choose a
regularization scheme: at order $g^0$, i.e. at the tree level, the
singularity of the two-point function needs to be subtracted and this
leads to the introduction of arbitrary contact terms. At higher
orders in $g$ (we have computed the two-point functions explicitly up
to the order $g^4$) divergences cancel out and as a consequence,
the local contact terms are determined
{\em unambiguously}. Thus the
contact terms in the two-point correlators are of the form
\beq
[a+f(g^2,N)] \, \pa^p \d^{(4)}(x-y)
\label{contact}
\eeq
with $a$ the arbitrary finite constant of the subtraction and $p$ an integer
depending on the dimensions
of the operators. The function $f(g^2,N)$
is in principle exactly computable order by order
in $g^2$ for any finite $N$, such that $f(g^2=0,N)=0$.

According to the AdS/CFT prescription one should establish a
correspondence between these terms and corresponding
ones in the supergravity sector.
There ambiguities arise due to the fact that the theory suffers from
infrared divergences when approaching the
boundary of the AdS space.
Within the holographic viewpoint proposed in \cite{adscft}, the
supergravity effective action,
evaluated on a solution of the equations of motion with prescribed
boundary conditions, becomes the
generating functional for the conformal field theory in the large $N$
limit.
The bulk fields $\phi$
evaluated at the boundary  act as source terms for composite operators
${\cal{O}}$ of the Yang-Mills theory
\beq
\langle e^{\int d^4x {\cal O} \phi_0}\rangle_{\rm CFT} ~=~
e^{S[\phi_0]}
\eeq
with the boundary action given by
\beq
S[\phi_0]= \int ~d^4x~d^4y~ \phi_0(x) \left[ \frac{1}{(x-y)^{2\D}}+
b \, (\pa^2)^{\D-2} \d^{(4)}(x-y)
\right]\phi_0(y)
\label{effaction}
\eeq
The local term, which acts as a regulator at
short distances in (\ref{effaction}), has an infrared origin in the 5d AdS
supergravity expanded
near the boundary \cite{UVIR, CS, HSS}.
The coefficient $b$ is a function of a mass scale parameter
identified with the inverse of the IR 5d cutoff \cite{SWPP}.

Now, comparing eq. (\ref{contact}) at small coupling with
eq. (\ref{effaction})
one should find that in the large $N$ limit, with the 't Hooft coupling
$g^2 N$ fixed but large
\beq
a+f(g^2,N)\rightarrow b_f
\label{strongequiv}
\eeq
where $b_f$ is the arbitrary finite part of the subtraction in
(\ref{effaction}).
In the context of our specific calculation we will show how this
identification can be realized.

We present this paper as a sequel to the one referred in \cite{us},
since the techniques used here to perform the perturbative calculations of the
correlators are the same as the ones used in \cite{us}.
Therefore in order to avoid lengthy repetitions in the main text, and at the
same time to make this paper self-contained, we have recollected
the main formulas and the rules of the game in the Appendices. In the
next Section we simply remind the reader which are the quantities we
have to focus on.
Then in Section 3 we enter
directly {\em in medias res}: we present the tree--level and the order
$g^2$ result for the $<{\rm Tr} (\Phi^1)^k~ {\rm Tr}
(\bar{\Phi}^1)^k>$
correlators. In Section 4 the order $g^4$ contributions are
considered. For the relevant diagrams we compute the colour
structure factors and the momentum integrals that one obtains after
completion of the $D$-algebra. The various, complicated contributions
give rise in the end to a complete cancellation for any finite $N$
\cite{LMRS}. This is in agreement with previous results \cite{DFS, HSW}.
In Section 5 we  show that the nonrenormalization properties we have
found
for the $SU(N)$ gauge group, are actually valid for general
 groups. Finally in Section 6 we
concentrate on the evaluation of the contact terms
and study what they correspond to in the supergravity sector.

\section{Two-point functions of chiral operators}

Our goal is to compute two-point
correlators for ${\cal{N}}=4$ supersymmetric $SU(N)$ Yang-Mills
theory, perturbatively in ${\cal N}=1$ superspace.
The operators under consideration are the chiral primary operators
in the $(0,k,0)$ representation of the $SU(4$) $R$--symmetry group. In a
${\cal N}=1$
superspace description of the  theory (see Appendix A for
details) they are given by
${\cal{O}}={\rm Tr} (\Phi^{\{i_1}\Phi^{i_2} \dots \Phi^{i_k\} })$,
with flavor indices on the superfields symmetrized and traceless. In
fact, in order to simplify matters as in Refs. \cite{DFS}, \cite{us},
we consider the $SU(3)$ highest weight superfield $\Phi^1$ and compute
$<{\rm Tr}(\Phi^1)^k{\rm Tr}(\Phib^1)^k>$.
In this way the flavor
combinatorics is avoided. At the same time we do not lose in generality
since the $SU(3)$ transformations, which are invariances of the theory,
allow to reconstruct
all the other primary chiral correlators from the one above.

The general strategy we have adopted for performing
the actual calculation is outlined in Appendix A.
Quite generally, at non--coincident points
we can write the two-point function as
\beq
<{\rm Tr}(\Phi^1)^k(z_1){\rm Tr}(\Phib^1)^k(z_2)>=
\frac{F(g^2,N)}{(x_1-x_2)^{2k}}\d^{(4)}(\theta_1-\theta_2)
\label{twopointtext}
\eeq
where $z \equiv (x,\theta, \bar{\theta})$.
Away from short distance singularities, the $x$-dependence
of the result is fixed by the conformal invariance of the theory,
and $F(g^2,N)$ is the function that we want to determine
perturbatively in $g^2$. As we have done in Ref. \cite{us} we compute
loop integrals in momentum space and
use dimensional regularization and minimal subtraction scheme to
treat ultraviolet divergences.
In $n$ dimensions, with $n=4-2\e$, the Fourier transform of a power factor
$(x_1-x_2)^{-2\n}$ is given by
\bea
\frac{1}{(x^2)^\n}=
2^{n-2\n}\pi^{\frac{n}{2}}\frac{\G(\frac{n}{2}-\n)}{\G(\n)}
\int \frac{d^n p}{(2\pi)^n} ~\frac{e^{-ipx}}{(p^2)^{\frac{n}{2}-\n}}
\qquad && \n \not= 0, -1, -2, \cdots \nonumber \\
&& \n \not= \frac{n}{2}, \frac{n}{2} +1, \cdots
\label{fourier}
\ena
The main advantage of such an approach is due to the fact
that the $x$--space structure  in (\ref{twopointtext}),
with a non--vanishing contribution to $F(g^2,N)$, can be obtained
simply by looking at the contributions that behave
like $1/\e$ from the singular factor
$\G(\frac{n}{2}-\n)=\G(-\n+2-\e)$, $\n \geq 2$ in (\ref{fourier}).
By analytic continuation one can write the general identity
\beq
\int \frac{d^n p}{(2\pi)^n} ~\frac{e^{-ipx}}{(p^2)^{2-k + \a \e}}
~=~ \frac{2^{2k-4}}{\pi^2} (-1)^k (k-1)! (k-2)! \a \,
\frac{\e}{(x^2)^{k - (\a+1)\e}} [ 1+ {\cal O}(\e) ]
\label{basicformula}
\eeq
Once the UV divergent terms are determined at a given
order in $g$, one can reconstruct the complete answer
using (\ref{basicformula}).

The UV divergent terms have been computed using
the method proposed in \cite{kazakov} and various techniques presented
in \cite{CT,russians}.
Infrared divergences have not been considered since the theory we are
dealing with is conformally invariant.

Finally we emphasize that finite momentum space
contributions to the correlators
correspond in $x$--space to terms proportional
to $\e$. These are the terms which give rise to contact terms \cite{CT}.
We will come back to this point in Section 5.

\section{At tree-level and order $g^2$ }

At tree--level, the 2--point correlation function
$<{\rm Tr}(\Phi^1)^k(z_1){\rm Tr}(\Phib^1)^k(z_2)>$ is given by the diagram
in Fig. 1. The colour structure
\beq
{\rm Tr}(T_{a_1} \cdots T_{a_k}) \, \sum_{\s} {\rm Tr} (T_{a_{\s(1)}}
\cdots T_{a_{\s(k)}} )
\label{treecolour}
\eeq
which includes all possible permutations $\s$ in the contractions
of the scalar lines, is a $k$--degree polynomial in $N$. In a double line
representation for the colour indices, the $N$-leading power term
corresponds to
the planar double line graph associated to the diagram in Fig. 1,
whereas nonplanar graphs give rise to subleading contributions.

\vskip 18pt
\noindent
\begin{minipage}{\textwidth}
\begin{center}
\includegraphics[width=0.55\textwidth]{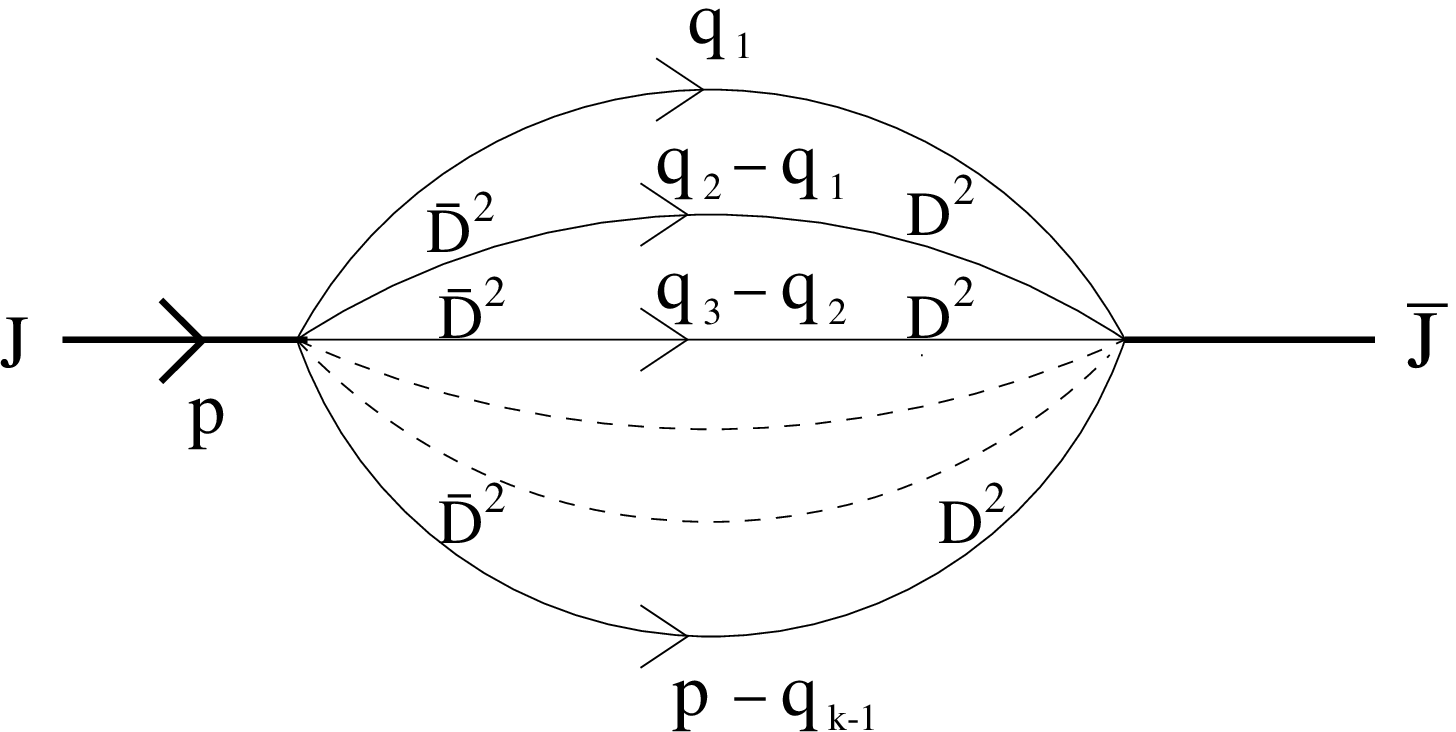}
\end{center}
\begin{center}
{\small{ Figure 1:
tree--level contribution to $<{\rm Tr}(\Phi_1)^k {\rm Tr}(\Phib_1)^k>$}}
\end{center}
\end{minipage}

\vskip 20pt

Now using the result (\ref{b5}) for the
momentum integrals to leading
order in the $\e$ expansion, we obtain
\beq
\frac{1}{\e}  \left[ \frac{1}{(4\pi)^2} \right]^{k-1} \,
\frac{(-1)^k}{[(k-1)!]^2} \, (p^2)^{k-2-(k-1)\e} \,
{\rm Tr}(T_{a_1} \cdots T_{a_k}) \, \sum_{\s} {\rm Tr} (T_{a_{\s(1)}}
\cdots T_{a_{\s(k)}} ) \, \d^{(4)}(\theta_1 -\theta_2)
\label{trelevel}
\eeq

In $x$--space, using eq. (\ref{basicformula}), the result can be rewritten
as
\bea
&& <{\rm Tr}(\Phi^1)^k(z_1){\rm Tr}(\Phib^1)^k(z_2)>_0  \nonumber \\
&&~~~~~~~~~~\nonumber \\
&&=~\left(\frac{1}{4\pi^2} \right)^k \,
{\rm Tr}(T_{a_1} \cdots T_{a_k}) \, \sum_{\s} {\rm Tr} (T_{a_{\s(1)}}
\cdots T_{a_{\s(k)}} )
\, \frac{1}{(x_1-x_2)^{2k}} \d^{(4)}(\theta_1 -\theta_2)
\label{corrfunct}
\ena

\vskip 20pt
At order $g^2$ the only contribution to the correlator
is given by the diagram in Fig. 2, with the insertion of a vector line.

\vskip 18pt
\noindent
\begin{minipage}{\textwidth}
\begin{center}
\includegraphics[width=0.45\textwidth]{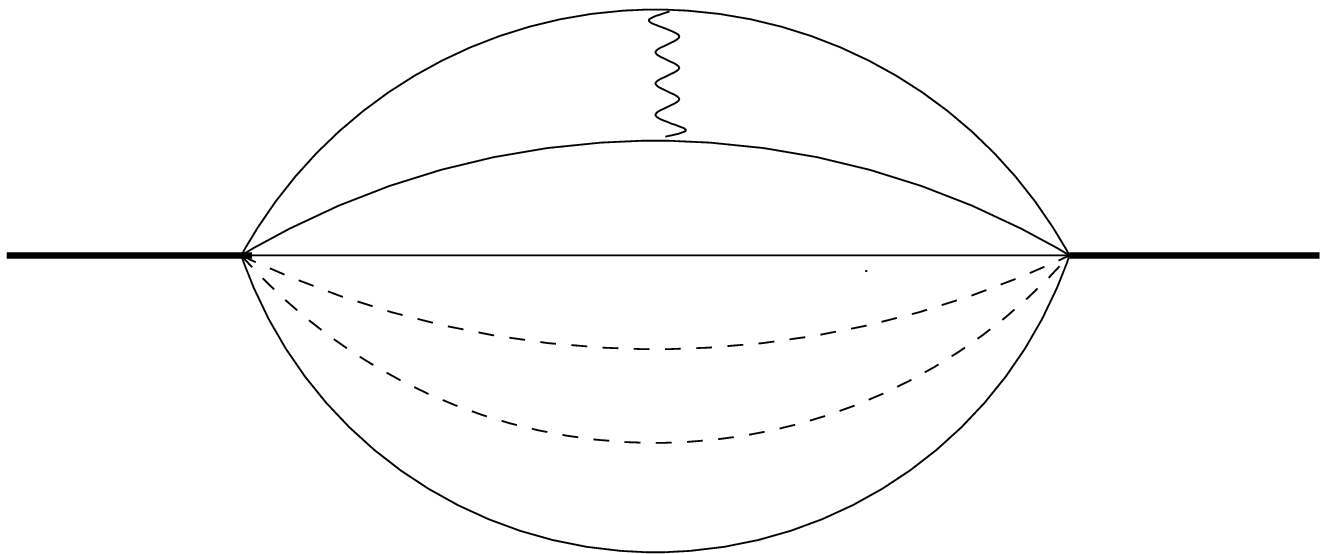}
\end{center}
\begin{center}
{\small{ Figure 2:
$g^2$--order contribution to $<{\rm Tr}(\Phi_1)^k {\rm Tr}(\Phib_1)^k>$}}
\end{center}
\end{minipage}

\vskip 20pt

From the two internal vertices $V_1$ in (A.6) we obtain the
colour structure
$f_{amb} f_{a' m b'}$ which contracted with
the colour matrices associated to the rest of the diagram gives
\beq
- {\rm Tr}(T_{a_1} \cdots T_{a_k}) \, \sum_{\s} \sum_{i \not= j}
{\rm Tr} (T_{a_{\s(1)}} \cdots [ T_{a_{\s(i)}}, T_m] \cdots
[T_{a_{\s(j)}}, T_m] \cdots T_{a_{\s(k)}} )
\label{loopcolour0}
\eeq
Here the sum is over all possible permutations of the external lines
and eq. (\ref{c1}) has been used. The previous expression
can be simplified \cite{DFS} by noticing that for any set of matrices
$M_j$, $j = 1, \cdots , n$, and any matrix $P$ the following identity holds
\beq
\S_{i=1}^n {\rm Tr} (M_1 \cdots [M_i,P] \cdots M_n) ~=~ 0
\label{generic}
\eeq
Thus (\ref{loopcolour0}) can be written as
\beq
{\rm Tr}(T_{a_1} \cdots T_{a_k}) \, \sum_{\s} \sum_{i=1}^k
{\rm Tr} (T_{a_{\s(1)}} \cdots [[ T_{a_{\s(i)}}, T_m], T_m] \cdots
T_{a_{\s(k)}} )
\eeq
which, using the identity (\ref{c2}), can be reduced further to the expression
in (\ref{treecolour}) up to a factor $2 N$.
Including the various factors from vertices and propagators, we finally
have
\beq
(g^2 N) \, k \,
{\rm Tr}(T_{a_1} \cdots T_{a_k}) \, \sum_{\s} {\rm Tr} (T_{a_{\s(1)}}
\cdots T_{a_{\s(k)}} )
\label{loopcolour1}
\eeq
The momentum integral associated to the graph after completion of the
$D$-algebra,
is evaluated in (\ref{b6}).
The final result (here we reinstate a factor $\frac{1}{(4\pi)^2}$
for each loop) is
\beq
12 \z(3) \, (g^2N) \, \left[ \frac{1}{(4\pi)^2} \right]^k \frac{(-1)^{k-1}
(k-1)}{[(k-1)!]^2} \, (p^2)^{k-2-k\e} \,
{\rm Tr} (T_{a_1} \cdots T_{a_k}) \, \sum_{\s} {\rm Tr} (T_{a_{\s(1)}}
\cdots T_{a_{\s(k)}} )
\label{g2level}
\eeq
In the limit $\e \rightarrow 0$ this expression is finite, therefore
it does
not contribute to the correlation function at separate points. It gives
rise to a contact term which
will be discussed in detail in Section 5.

\section{At order $g^4$}

The relevant diagrams, drawn in single line representation, are
shown in Fig. 3 and 4.
They group themselves into planar ($A$) and nonplanar
($B$) ones. If one were to use a double line representation for the colour
indices from $T^a_{ij}$, then the single-line planar graphs ($A$) would give
rise to two distinct types of graphs, double-line planar graphs ($A_1$)
and nonplanar double-line graphs ($A_2$).
Now from the $A_1$ graphs the colour
structure would produce leading $N$ contributions, while from
the $A_2$ graphs subleading contributions would
arise. Obviously the single-line type ($B$) diagrams could only give
structures subleading in $N$  since their double-line representation
would be necessarily nonplanar.

\vskip 18pt
\noindent
\begin{minipage}{\textwidth}
\begin{center}
\includegraphics[width=0.70\textwidth]{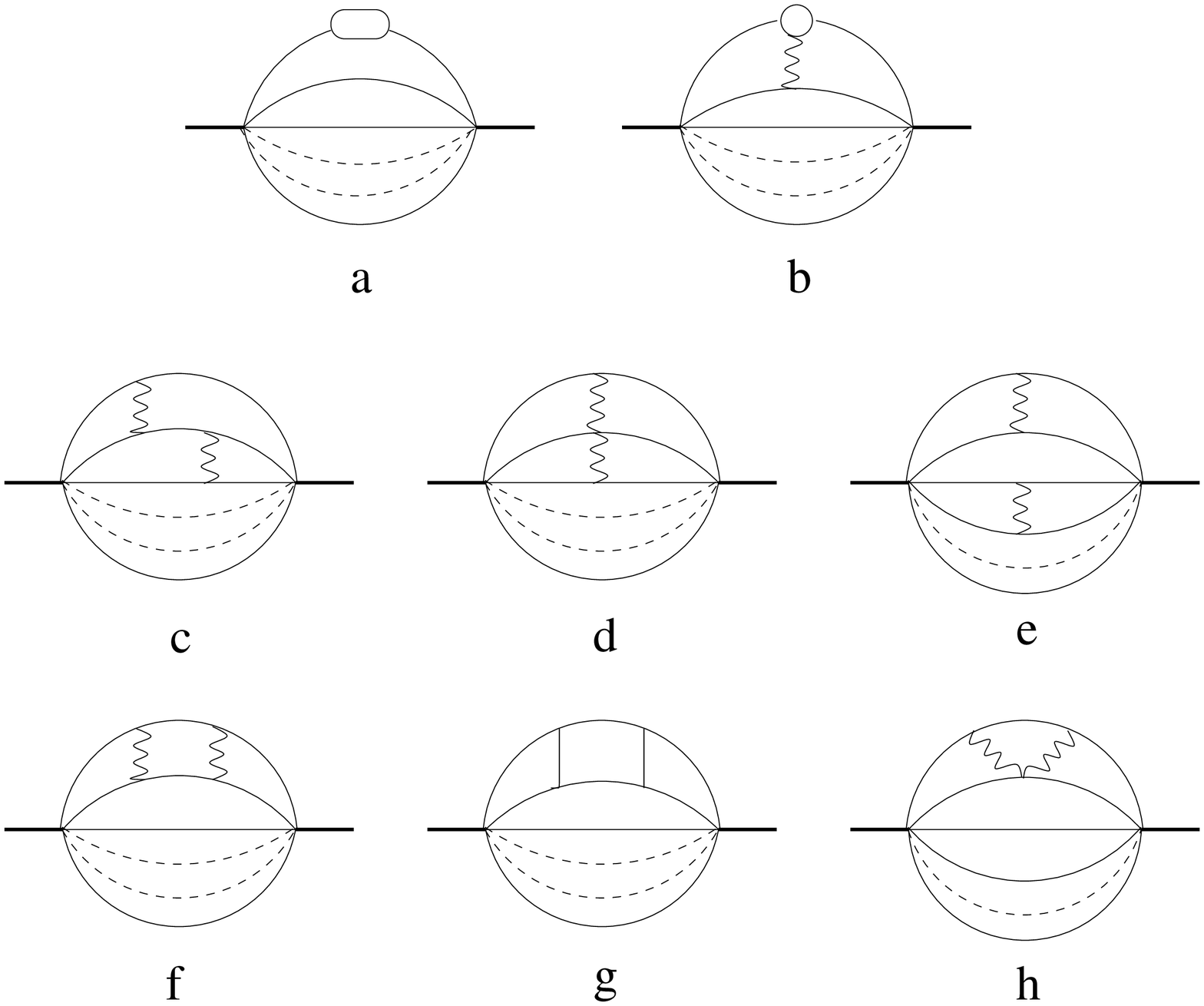}
\end{center}
\begin{center}
{\small{ Figure 3:
planar $g^4$--order contribution to $<{\rm Tr}(\Phi_1)^k
{\rm Tr}(\Phib_1)^k>$}}
\end{center}
\end{minipage}

\vskip 20pt

\vskip 18pt
\noindent
\begin{minipage}{\textwidth}
\begin{center}
\includegraphics[width=0.70\textwidth]{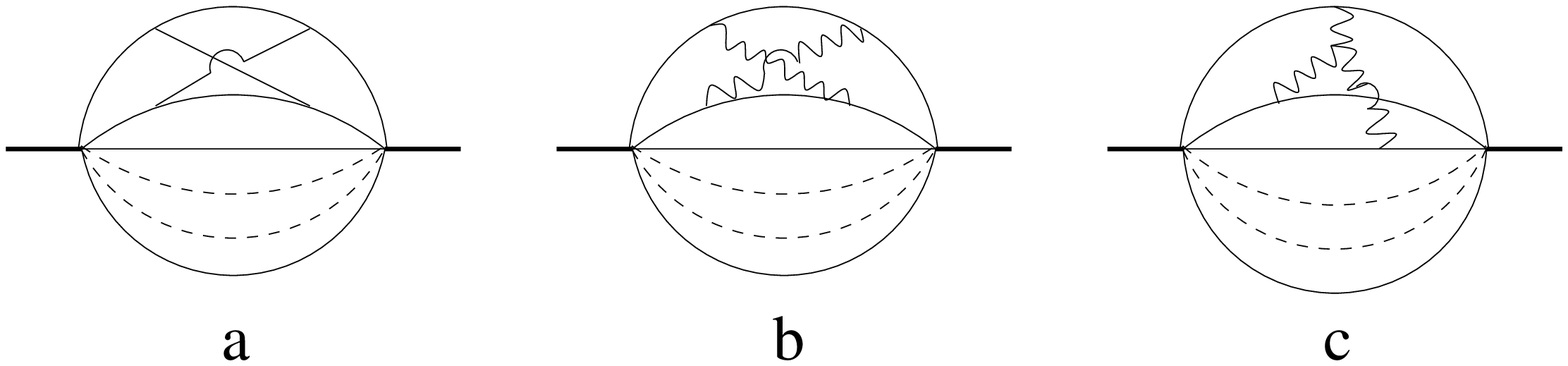}
\end{center}
\begin{center}
{\small{ Figure 4:
nonplanar $g^4$--order contribution to $<{\rm Tr}(\Phi_1)^k
{\rm Tr}(\Phib_1)^k>$}}
\end{center}
\end{minipage}

\vskip 20pt

For the correlators computed in \cite{us}, i.e. for the $k=3$ case,
the colour factor of the diagrams in the class ($B$) turned out to be
identically zero. All the graphs in the class ($A$) gave rise to the
same overall combinatorics, so that the final cancellation occurred
as a cancellation among the various terms produced by the momentum
integrations.

In the present case, $k>3$, we will show that the situation is much
more complicated and highly non trivial. All types of
diagrams mentioned above have non vanishing colour structures.
The diagrams ($A$) give contributions proportional to
the colour structure (\ref{treecolour}) which,
after multiplication by the momentum integrals,
sum up to zero. However in addition, from the diagrams in Fig. $3c$
and $3d$ a new subleading structure arises which is of the same kind as
the one from diagrams ($B$). Again, a nontrivial cancellation occurs
among the diagrams in Fig. $3c$, $3d$ and ($B$) due to the special
structure of their momentum integrals.

Now for each of diagram we present  the evaluation of the colour factor,
while we list the corresponding momentum integrals in Appendix B.
The final results are summarized
at the end of the Section where the actual cancellation is discussed.

In Fig. $3a$ we have the insertion of a two--loop propagator
correction \cite{GSR,us}
\bea
&&-2g^4~N^2~\Phib^i_a(p,\theta)~ \Phi^i_a(-p,\theta)~p^2~
\int \frac{d^n q~d^nk}{k^2q^2(k-q)^2(k-p)^2(p-q)^2}\nonumber\\
&&~~~~~~~~~~~=-2g^4~N^2~ \Phib^i_a(p,\theta)~
\Phi^i_a(-p,\theta)~\frac{1}{(p^2)^{2\e}}
[6\zeta(3)+{\cal O}(\e)]
\label{2prop}
\eea
In this case the colour structure is easy to compute. Inserting
all the coefficients from vertices, propagators and combinatorics we obtain
\beq
12 \z(3) (g^2N)^2 \, k \,
{\rm Tr}(T_{a_1} \cdots T_{a_k}) \, \sum_{\s} {\rm Tr} (T_{a_{\s(1)}}
\cdots T_{a_{\s(k)}} )
\label{3acolour}
\eeq
The momentum integral emerging after the $D$--algebra is
given in eq. (\ref{b7}).

We now consider the diagram $3b$  where the ${\cal O}(g^2)$ effective vertex
\cite{GSR} appears
\bea
&&\frac{g^3}{4}~ N ~if^{abc} ~\Phib^i_a (q,\theta)\Phi^i_b(-p,\theta)
\left( 4D^\a\Db^2D_\a \right. \nonumber\\
&&~~~~~~~~~~~~~~~~~~~~\left.+(p+q)^{\a\ad}
[D_\a,\Db_\ad ]\right) V_c (p-q,\theta)~\int \frac{d^nk}{k^2(k-p)^2(k-q)^2}
\label{3vertex}
\eea
Performing the $D$--algebra one easily realizes that only the first
term in (\ref{3vertex}) gives rise to potentially divergent
loop integrals. The second term produces finite contributions
which might generate  order $g^4$ contact terms.  We concentrate
on the $D^\a\Db^2D_\a$ term, since we will not compute
contact contributions to  $g^4$ order.
 The colour structure for this diagram
is computed following the same procedure as in the case of the
diagram  in Fig. 2 (see eqs. (\ref{loopcolour0}--\ref{loopcolour1})).
Inserting all the various factors we obtain
\beq
-4 (g^2 N)^2  \, k
{\rm Tr}(T_{a_1} \cdots T_{a_k}) \, \sum_{\s} {\rm Tr} (T_{a_{\s(1)}}
\cdots T_{a_{\s(k)}} )
\label{3bcolour}
\eeq
The corresponding momentum integral is given in eq. (\ref{b8}).

We now turn to the discussion of the colour structure for the diagram
in Fig. $3c$. The insertion of the two vector lines gives rise to
the structure
$f_{a_1mb_1} f_{a_2 m n} f_{n p b_2} f_{a_3 p b_3}$. The contraction
with scalar lines from the external vertices takes into account all possible
permutations. Using eq. (\ref{c1}) we can write
\beq
{\rm Tr}(T_{a_1} \cdots T_{a_k}) \, \sum_{\s} \sum_{i \not= j \not= l}
{\rm Tr} (T_{a_{\s(1)}} \cdots [T_{a_{\s(i)}} ,T_m] \cdots
[[T_{a_{\s(j)}} ,T_m], T_n] \cdots [T_{a_{\s(l)}} ,T_n]
\cdots T_{a_{\s(k)}} )
\label{intermediate}
\eeq
This expression can be manipulated by making use of
the identity (\ref{generic}) and it can be written
as the sum of two terms
\bea
&& - {\rm Tr}(T_{a_1} \cdots T_{a_k}) \, \sum_{\s} \sum_{j \not= l}
\left\{ {\rm Tr} (T_{a_{\s(1)}} \cdots
[[[T_{a_{\s(j)}} ,T_m], T_n],T_m] \cdots [T_{a_{\s(l)}} ,T_n]
\cdots T_{a_{\s(k)}} ) \right.  \nonumber \\
&&~~~~~~~~~~~\left. + {\rm Tr}(T_{a_{\s(1)}} \cdots
[[T_{a_{\s(j)}} ,T_m],T_n]
\cdots [[T_{a_{\s(l)}},T_n],T_m] \cdots T_{a_{\s(k)}} ) \right\}
\ena
The first term can be further reduced by using the identities (\ref{c2}),
(\ref{c3}) and (\ref{generic}) again. Inserting the factors from
combinatorics, vertices and propagators the final expression for the colour
structure can be written as the sum of a term leading in $N$ plus a
subleading contribution
\bea
&& g^4\, {\rm Tr}(T_{a_1} \cdots T_{a_k}) \,
\sum_{\s} \big\{ 2 k  N^2 {\rm Tr} (T_{a_{\s(1)}} \cdots
T_{a_{\s(k)}} ) \nonumber \\
&&~~~~~~~~~~~- \sum_{j \not= l} {\rm Tr}(T_{a_{\s(1)}} \cdots
[[T_{a_{\s(j)}} ,T_m],T_n]
\cdots [[T_{a_{\s(l)}},T_n],T_m] \cdots T_{a_{\s(k)}} ) \big\}
\label{3ccolour}
\ena
For $k=3$ the second term vanishes and the above expression reduces
to the one computed in Ref. \cite{us}. However for $k>3$ the subleading
term is nonzero, as shown in Appendix C explicitly for the $k=4$ case.
For this diagram the loop--integral result can be read in eq. (\ref{b9}).

The colour factor for the diagram in Fig. $3d$ can be computed by
exploiting the previous results. In fact, from the internal vertices
$V_1$ and $V_3$ (see eq. (A.6)) we have the structure $f_{a_1 m b_1}
f_{p b_2 n} f_{m a_2 n} f_{a_3 p b_3}$ which is identical to the one
from the diagram $3c$. Performing all the contractions and taking into
account the coefficients from combinatorics, vertices and propagators
we finally obtain
\bea
&& \frac{g^4}{2}\, {\rm Tr}(T_{a_1} \cdots T_{a_k}) \,
\sum_{\s} \Big\{ 2 k N^2 {\rm Tr} (T_{a_{\s(1)}} \cdots
T_{a_{\s(k)}} )  \nonumber \\
&&~~~~~~~~~ - \sum_{j \not= l} {\rm Tr}(T_{a_{\s(1)}} \cdots
[[T_{a_{\s(j)}} ,T_m],T_n]
\cdots [[T_{a_{\s(l)}},T_n],T_m] \cdots T_{a_{\s(k)}} ) \Big\}
\label{3dcolour}
\ena
The momentum integral for this graph is given in eq. (\ref{b10}).

The last potential contributions from diagrams of type $(A)$ are shown
in Figures $3e-3h$.
The loop integral resulting from the D--algebra on the diagram $3e$ given
in eq. (\ref{b11}), is ${\cal O}(\e)$ and then it does not contribute to the
correlation function. The diagrams in Figures $3f$, $3g$ and $3h$, exactly as
for $k=3$, only contribute to finite terms.

We now study the nonplanar diagrams ($B$) in Figure 4.\\
The combinatorial and colour factors, and the loop integrals
for the two diagrams $4a$ and $4b$ are identical. Thus we
concentrate on one of them, e.g. $4b$. From the internal $V_1$
vertices with vector lines contracted as in figure, the colour factor
which arises is $f_{a_1 m n} f_{n p b_1} f_{a_2 p q} f_{q m b_2}$.
When connected with the rest of the diagram it gives
\beq
\frac{g^4}{2}
\sum_{\s} \sum_{j \not= l} {\rm Tr}(T_{a_{\s(1)}} \cdots
[[T_{a_{\s(j)}} ,T_m],T_n]
\cdots [[T_{a_{\s(l)}},T_n],T_m] \cdots T_{a_{\s(k)}} )
\label{4acolour}
\eeq
where combinatorial factors have been included.
We note that the previous expression has the same form as the subleading
contribution already present in the diagrams $3c$ and $3d$ (see eqs.
(\ref{3ccolour}, \ref{3dcolour})). As previously mentioned and proven in
Appendix C, this term is
zero for $k=3$ but in general nonvanishing when $k >3$.
The corresponding loop diagram arising after the D--algebra is given in
eq. (\ref{b8}).

The last nonplanar graph to be considered is the one in Fig. $4c$. As in
the $k=3$ case \cite{us}, its colour coefficient vanishes.
A simple way to prove this is to notice that from the internal vertices
one obtains
\beq
f_{a_1 m b_1} f_{a_2 n b_2}
f_{a_3 p b_3} f_{m n p}
\eeq
which is antisymmetric under
the exchange $a_1 \leftrightarrow a_2$ and
$b_1 \leftrightarrow b_2$.  When multiplied by
all possible permutations of colour matrices from the external vertices
it gives a zero result.

\vskip 20pt
We now collect all the divergent contributions
we have computed at order $g^4$.
For each diagram we list the final result obtained as a product
of the colour and combinatorial factors times the
results from momentum integrations.
We define
\bea
  {\cal  P}_k && \equiv~ {\rm Tr}(T_{a_1} \cdots
T_{a_k}) \sum_{\s} {\rm Tr}(T_{a_{\s(1)}} \cdots T_{a_{\s(k)}})
\\
{\cal Q}_k && \equiv~
{\rm Tr}(T_{a_1} \cdots T_{a_k}) \sum_{\s} \sum_{j \not= l}
{\rm Tr}(T_{a_{\s(1)}} \cdots [[T_{a_{\s(j)}}, T_m],T_n] \cdots
[[T_{a_{\s(l)}}, T_n], T_m] \cdots T_{a_{\s(k)}}) \nonumber
\label{T}
\ena
Factorizing an overall common coefficient
\beq
\frac{1}{\e} \, (g^2N)^2 \, \left[ \frac{1}{(4\pi)^2} \right]^{k+1}
\frac{(-1)^k (k-1)}{[(k-1)!]^2 (k+1)} (p^2)^{k-2-(k+1)\e}
\label{common}
\eeq
the various contributions are

\begin{itemize}
\item
Diagram 3a:
\beq
12k \z(3) \,   {\cal  P}_k
\label{r1}
\eeq
\item
Diagram $3b$:
\beq
-24k \z(3) \,   {\cal  P}_k
\label{r2}
\eeq
\item
Diagram $3c$:
\beq
[12k \z(3) - 40k \z(5)] \,   {\cal  P}_k
+ \frac{1}{N^2} [ 20\z(5) - 6\z(3)] {\cal Q}_k
\label{r3}
\eeq
\item
Diagram $3d$:
\beq
40k \z(5) \,   {\cal  P}_k
- \frac{1}{N^2} 20\z(5) {\cal Q}_k
\label{r4}
\eeq

\item
Diagrams $4a + 4b$:
\beq
\frac{1}{N^2} 6\z(3) \, {\cal Q}_k
\label{r5}
\eeq
\end{itemize}
The leading structure  $  {\cal  P}_k$ appears
in the diagrams $3a$ and $3b$
with a coefficient proportional to the $\z(3)$ Riemann function,
whereas in the
diagram $3d$ it is proportional to $\z(5)$. The same pattern repeats itself for
the non-leading structure ${\cal Q}_k$:
in the diagrams $4a$ and $4b$ it is multiplied
by $\z(3)$, whereas in the diagram $3d$ it is proportional to  $\z(5)$.
Both structures appear in the diagram $3c$  and they contribute with
such a coefficient to cancel the rest of the divergent terms.
In conclusion, only finite
contributions survive. As emphasized at the beginning this amounts to say
that the correlation function
is not renormalized at order $g^4$, up to contact terms.

We stress that the complete cancellation of divergent contributions
for general $k$ occurs for {\em any} finite $N$.

The results in (\ref{r1}-\ref{r5}) if restricted to
$k=3$, reproduce the results discussed
in Ref. \cite{us}. In this case the subleading structure
${\cal Q}_k$ vanishes, so that the proof of nonrenormalization
for the correlator with general $k$ cannot be implemented from
the $k=3$ example.

\section{General gauge groups}

Here we want to show that the nonrenormalization properties of the
correlators proven in the two previous Sections for the $SU(N)$ super
Yang-Mills theory, actually hold for general simple gauge groups.
To this end it is sufficient to realize that, using (\ref{ff}) and
(\ref{fff}) which are valid for any group, we can generalize the
identities in (\ref{c2}) and (\ref{c3}) to the following ones
\beq
[[T_a, T_m], T_m] ~=~ k_1 \, T_a
\label{c2g}
\eeq
and
\beq
[[[T_a, T_m], T_n], T_m] ~=~ \frac{k_1}{2} \, [T_a, T_n].
\label{c3g}
\eeq
At the order $g^2$ this enables us to replace $2N$ with $k_1$
in (\ref{loopcolour1}) and
(\ref{g2level}),  thus obtaining a result which depends only
on the Casimir of the adjoint representation of the gauge group.
In any event to this order the
momentum integral gives a finite result which does not affect the
correlator at non coincident points.

In the same way one can analyze the general situation at the next perturbative
 $g^4$ order. For the propagator and vertex insertions which appear
 in the graphs of Fig. $3a$, $3b$ we use the result as in Ref.
 \cite{GSR}, i.e. we set $2N\rightarrow k_1$ in (\ref{2prop}) and in
 (\ref{3vertex}).
For the rest of the diagrams we
 substitute (\ref{c2}), (\ref{c3}) with (\ref{c2g}), (\ref{c3g})
respectively. Once this operation is performed consistently
everywhere in the various formulas of Section 4 the final, complete
cancellation of all the corrections is achieved following exactly the
same pattern and the same steps as in the $SU(N)$ case.

\section{Contact terms in the AdS/CFT correspondence}

Now we wish to focus our attention on the
two-point correlators for $SU(N)$
when the two points approach each other. In the
limit $x_1 \rightarrow x_2$ the expression in (\ref{twopointtext})
becomes singular and needs to be regulated. Within the dimensional
regularization approach we have adopted,
this short-distance singularity is signaled by $1/\e$ poles, according to
the general identity
\beq
\frac{1}{(x^2)^{k-k\e}} \sim
-\frac{\pi^2}  {2^{2k-4}(k-1)! (k-2)!} ~\frac{1}{\e}~ (\pa^2)^{k-2}
\d^{(n)}(x) \qquad {\rm for} ~~ \e \to 0
\label{contactdeltak}
\eeq
For the two--point function of the operator
${\rm Tr}(\Phi^1)^k$ these short-distance UV divergences manifest
themselves already at tree--level (see (\ref{corrfunct})).
In order to obtain a well defined function at coincident points we
perform a subtraction and define in configuration space
\bea
&& <{\rm Tr}(\Phi^1)^k (z_1) {\rm Tr}(\Phib^1)^k (z_2)>_{\rm reg} \equiv
\d^{(4)}(\theta_1 -\theta_2) \times \nonumber\\
&& ~~~~~~~~~~\nonumber \\
&& \lim_{\e \to 0} \, \frac{  {\cal  P}_k}{(4\pi)^{2k}}
\, \left[ \frac{1}{(x^2)^{k-k\e}}
~+~ (-1)^k (\m^2)^{(1-k)\e} \frac{2^{4-2k} \pi^2}{[(k-1)!]^2}
\left( \frac{1}{\e} +\g \right) (\pa^2)^{k-2} \d^{(n)}(x_1-x_2)\right]
\nonumber \\
&& ~~~~~~~~~~~
\label{regfunct}
\ena
where $  {\cal  P}_k$ has been defined in (4.12)
and $\m$ is the mass scale of dimensional regularization.
The coefficient $\g$ corresponds to an arbitrary finite subtraction
and generates a scheme dependent finite contact term.

Performing explicitly the $\e \to 0$ limit in (\ref{regfunct})
a residual dependence on the
mass scale survives from the (divergent) counter--contact term. Therefore,
the subtraction of the infinity at tree--level necessarily introduces
a mass scale in the regularized correlation function which breaks
conformal invariance \cite{peske}. On the other hand, the scheme
dependent finite term proportional to $\g$  is independent of
$\m$ and it does not affect conformal invariance.

Now we discuss the appearance of this type of terms in the
perturbative computation of
the two--point correlator.
We have performed our loop calculations in momentum space, with the
Fourier transformation given in
the basic formula ({\ref{basicformula}). Using ({\ref{basicformula})
and the general identity in (\ref{contactdeltak}) we can write
\beq
\lim_{\e \to 0} \int \frac{d^n p}{(2\pi)^n}
\frac{e^{-ipx}}{(p^2)^{\frac{n}{2}-k+\a \e}}
= (-1)^{k+1}\, \a \, (\pa^2)^{k-2} \d^{(4)}(x)
\label{delta}
\eeq
i.e. any finite contribution in momentum space gives rise to finite contact
terms in the correlation functions.

In general, the presence of contact terms at any loop order
is related to the UV regularization
and renormalization procedure. Different subtraction conditions
correspond to different finite counterterms which eventually contribute to
contact terms in the correlation functions of the theory.
 In $n =4-2\e$ dimensions, one has to choose a particular
regularization for evaluating the integrals
\beq
\int d^4p \, f(p) \rightarrow G(\e) \int d^np \, f(p)
\label{gscheme}
\eeq
where $G(\e)$ is a regular function\footnote{A convenient choice
is $G(\e) = (4\pi)^{-\e}
\G(1-\e)$ (G--scheme \cite{russians})
which cancels irrelevant terms proportional to
the Euler constant, $\log{4\pi}$
and $\z(2)$ Riemann function in the $\e$--expansion of $\G$ functions which
appear in the calculation of multiloop integrals.}
near $\e=0$, with $G(0) =1$.
A given prescription has to be used in the computation
of both the effective action and physical quantities like correlation
functions or scattering amplitudes.
By expanding $G(\e)$ in powers of $\e$ one can easily realize that
in multiloop integrals with only simple pole divergences the
coefficients of the divergent terms do not
depend on the particular choice of the $G$--function, whereas they might
depend on the regularization scheme in multiloop integrals with higher
order divergences.
In any case, a scheme dependence is always present in the finite part of
{\em any} divergent diagram.
It follows that in the evaluation of correlation functions, different choices
of the $G$--function, i.e. different
regularization prescriptions, give rise in general to different finite
quantum contact terms.
However, if a nonrenormalization theorem holds, then the contact terms
become independent of the
regularization prescription and they are {\em unambiguously} computable.

Indeed let us exemplify for simplicity the case in which
at a given loop order the perturbative contribution
to a correlation function contains at most a simple
pole $1/\e$ divergence
\beq
\left( \frac{a}{\e} + b + O(\e) \right) G(\e)
\eeq
By expanding
$G(\e) = 1 + c\e + \cdots$, in the limit $\e \to 0$ we obtain
the divergent scheme independent contribution $a/\e$ and the
finite term $(b + c a)$
which contains a scheme dependence through the coefficient $c$ from $G(\e)$.
However, if there is no perturbative renormalization, i.e. $a = 0$,
the finite term is uniquely determined.

In our specific case, we have shown that up to the $g^4$--order
the two--point function
$< {\rm Tr} (\Phi^1)^k (z_1){\rm Tr} (\Phib^1)^k(z_2)>$ is not perturbatively
renormalized and we are in the situation where the
finite contact terms are uniquely determined.
The term at order $g^2$ can be easily inferred from (\ref{g2level})
by using
the identity (\ref{delta}). In the limit $\e \to 0$ we find
\beq
12\z(3) (g^2N) \frac{k^2-1}{[(k-1)!]^2} \frac{  {\cal  P}_k}{(4\pi)^{2k}}
\, (\pa^2)^{k-2} \d^{(4)} (x_1-x_2) \d^{(4)}(\theta_1 -\theta_2)
\label{g2contact}
\eeq
The same procedure can be applied in order to compute the contact
term at order $g^4$. One should keep track of all finite
contributions  from the momentum integrals and then use
(\ref{delta}) to obtain the result in configuration space.
We have not performed the
explicit calculation but in general we expect to find a nonvanishing
result.

We note that, since these loop contact terms come from finite contributions
in the $\e$--expansion, in the four
dimensional limit no dependence on the scale $\m$ survives. Therefore
these terms do not affect the conformal invariance properties of the theory.
As discussed in Ref. \cite{peske}
the breaking of conformal invariance can only be ascribed to the
UV divergences in the correlation function.

If the non--renormalization theorem holds for any finite $N$ at all orders
in perturbation theory, one could determine unambiguously
the finite contact terms loop by loop, and generate a contribution of the form
\beq
[ a ~+~ f(g^2,N)] \, (\pa^2)^{k-2} \d^{(4)} (x_1-x_2)
\d^{(4)}(\theta_1 -\theta_2)
\label{contatto}
\eeq
where the arbitrary tree--level coefficient $a$ is given by
\beq
a \equiv \g (-1)^k
\frac{  {\cal  P}_k}{(4\pi)^{2k}} \frac{2^{4-2k}\pi^2}{[(k-1)!]^2}
\eeq
Our result shows that the complete answer for  the two--point functions
of the theory necessarily contains contact terms, unless we were to
choose a non-minimal subtraction at
short distances. This would amount to the
subtraction of a {\em finite} term with a coefficient $a=-f(g^2,N)$.

The SL(2,{\cal Z) invariance of the theory would require
$f(g^2,N)$ to be a modular function. However,
following the arguments in \cite{intril}, if loop contact terms
were present in the
final result they would necessarily break the $U(1)_Y$ invariance that
the theory inherits from the 5d supergravity. It might be possible to
establish a correspondence with $U(1)$--breaking terms in the
supergravity action \cite{green}.

According to the AdS/CFT conjecture, the generating functional of the
regularized correlation functions in the large $N$ limit
is the semiclassical 5d supergravity
action with IR divergences suitably subtracted \cite{peske}. At the
boundary, i.e. with the IR cut--off removed, it has the form
(\ref{effaction}).
In particular, the arbitrary finite contact term in (\ref{contatto})
corresponds to an arbitrary finite subtraction in 5d
\beq
S_c[\phi_0] ~=~ b_f \int d^4x \, \phi_0(x) (\pa^2)^{k-2} \phi_0(x)
\eeq
related
to the particular IR regularization scheme chosen for the supergravity action.
In other words, in the large $N$ limit with $g^2N \gg 1$
we should find
\beq
[ a + f(g^2,N)] \rightarrow b_f
\label{limit}
\eeq
If in this limit $f(g^2,N)$ is not vanishing, we might conclude that
either coupling dependent local terms appear in supergravity, or
one is forced to choose a particular subtraction scheme in the Yang--Mills
sector.

Finally we notice that perturbative contact terms in the two--point
functions give rise in general to coupling dependent contact--type
contributions in higher--point correlators, as well as enter the definition of
multitrace operators through the point--splitting regularization \cite{CS}.

\medskip

\section*{Acknowledgements}
\noindent We wish to thank Misha Bershadsky, Dan Freedman and
Michael Green for discussions and suggestions. \\
\noindent This work has been supported by the European Commission TMR
programme ERBFMRX-CT96-0045, in which S.P. and D.Z. are associated
to the University of Torino.

\newpage

\appendix

\section{The basic rules for the computation of the two-point correlators}

In ${\cal N}=1$ superspace the action  of ${\cal N}=4$ supersymmetric
Yang-Mills
theory can be written in terms of one real
vector
superfield $V$ and three chiral superfields $\Phi^i$
(we follow the notations in \cite{superspace})
\bea
S[J,\bar{J}]
&=&\int d^8z~ {\rm Tr}\left(e^{-gV} \Phib_i e^{gV} \Phi^i\right)+
\frac{1}{2g^2}
\int d^6z~ {\rm Tr} W^\a W_\a\nonumber\\
&&+\frac{ig}{3!} {\rm Tr} \int d^6z~ \e_{ijk} \Phi^i
[\Phi^j,\Phi^k]+\frac{ig}{3!} {\rm Tr} \int d^6\bar{z}~ \e_{ijk} \Phib^i
[\Phib^j,\Phib^k]\nonumber\\
&&+\int d^6z~ J {\cal O}+\int d^6\bar{z}~ \bar{J}\bar{{\cal O}}
\label{actionYM}
\eea
where $W_\a= i\Db^2(e^{-gV}D_\a e^{gV})$, and $V=V^aT^a$,
$\Phi_i=\Phi_i^a T^a$,
$T^a$ being $SU(N)$ matrices in the fundamental representation.
We have added to the classical
action source terms for the chiral primary operators generically denoted by
${\cal O}$.

We define the generating functional in Euclidean space
\beq
W[J,\bar{J}]=\int {\cal D}\Phi~{\cal D}\Phib~{\cal D}V~e^{S[J,\bar{J}]}
\label{genfunc}
\eeq
so that for ${\cal O} = {\rm Tr} (\Phi^1)^k$
the two-point function is given by
\beq
<{\rm Tr}(\Phi^1)^k(z_1){\rm Tr}(\Phib^1)^k(z_2)>=
\left. \frac{\d^2 W}{\d J(z_1)\d\bar{J}(z_2)}\right|_{J=\bar{J}=0}
\label{defcorr}
\eeq
where $z \equiv (x,\theta, \bar{\theta})$.
We use perturbation theory to evaluate the contributions to $W[J,\bar{J}]$
which are
quadratic in the sources, i.e.
\beq
W[J,\bar{J}]\rightarrow \int d^4x_1~d^4x_2~ d^4\theta~
J(x_1,\theta,\bar{\theta})
\frac{F(g^2,N)}{(x_1-x_2)^{2k}}\bar{J}(x_2,\theta,\bar{\theta})
\label{twopoint}
\eeq
The $x$-dependence
of the result is fixed by the conformal invariance of the theory,
and $F(g^2,N)$ is the function to be determined.

In order to obtain
the result in (\ref{twopoint}) one has to consider all the two-point
diagrams from $W[J,\bar{J}]$ with $J$ and $\bar{J}$ on the external
legs.
First one evaluates
all factors coming from combinatorics and colour
structures of a given diagram. Then one performs the
superspace $D$-algebra following standard techniques
(see for example \cite{superspace}), and
reduces the result to a multi-loop integral.

The quantization procedure of the classical
action in (\ref{actionYM}) requires the introduction of a gauge fixing
(we work in Feynman gauge) and corresponding ghost terms.
The ghost superfields only couple to the vector
multiplet and are not interesting for our calculation.
In momentum space we have the superfield propagators
\beq
<V^a V^b>= - \frac{\d^{ab}}{p^2}\qquad\qquad
<\Phi^a_i \Phib^b_j>=\d_{ij} \frac{\d^{ab}}{p^2}
\label{propagators}
\eeq
The vertices are read directly from the interaction terms in (\ref{actionYM}),
with additional $\Db^2$, $D^2$ factors for chiral, antichiral lines
respectively.
The ones that we need are the following
\bea
&&V_1=ig f_{abc}\d^{ij} \Phib^a_i V^b \Phi^c_j \qquad\qquad \qquad
V_2=-\frac{i}{2}g f_{abc}V^a \Db^2 D^\a V^b D_\a V^c \nonumber\\
&&~~~~~~~~~~~~~~~~~
V_3=\frac{g^2}{2}  \d^{ij} f_{adm} f_{bcm} V^aV^b \Phib^c_i
\Phi^d_j \\
&& V_4 = - \frac{g}{3!} \e^{ijk} f_{abc} \Phi_i^a \Phi_j^b \Phi_k^c
\qquad \qquad \qquad
\bar{V}_4 = - \frac{g}{3!} \e^{ijk} f_{abc} \Phib_i^a \Phib_j^b \Phib_k^c
\nonumber
\label{vertices}
\eea
All the calculations are performed in $n$ dimensions with $n=4-2\e$ and in
momentum space.
We have used the method of uniqueness \cite{kazakov} which
is particularly efficient
for the computation of massless Feynman integrals of a single variable.

\section{Relevant integrals in momentum space}

In this Appendix we list the relevant multiloop integrals which have been
used in the course of our calculation.

As described in Section 5,
in dimensional regularization $n =4-2\e$, one has to choose a particular
prescription for the regularized integrals.
In our case the integrals have at the most $1/\e$ divergences so
that they do not depend on the particular choice of
the $G$--function,
as far as divergent contributions are concerned.
In order to simplify the calculation, we forget about
$(2\pi)$ factors
at intermediate stages and reinsert at the end a $1/(4\pi)^2$
factor for each loop. Having this in mind we now list the
relevant integrals and their $\e$ expansion.

The basic integrals from which all our results can be deduced are the
following ones:\\
At one loop
\beq
I_1=\int  \frac{d^nk}{(k^2)^\a [(p-k)^2]^\b}=
\frac{\G(\a+\b-\frac{n}{2} )}{\G(\a)\G(\b)} \frac{\G(\frac{n}{2}-\b)\G(\frac{n}{2} -\a)}
{\G(n-\a-\b)} \frac{1}{(p^2)^{\a+\b-\frac{n}{2} }}
\label{b1}
\eeq
At two loops
\beq
I_2=\int \frac{d^nk~d^nl}{k^2l^2(k-l)^2(p-k)^2(p-l)^2}= \frac{1}{(p^2)^{1+2\e}}
[6\z(3)+{\cal O}(\e)]
\label{b2}
\eeq
At four loops
\beq
I_4=\int \frac{d^nk~d^nl~d^nr~d^ns}{k^2l^2(k-l)^2(r-k)^2(s-l)^2(r-s)^2
(p-r)^2(p-s)^2}
=\frac{1}{\e}\, \frac{1}{(p^2)^{4\e}}~[5\z(5)+{\cal O}(\e)]
\label{b3}
\eeq
and
\bea
&&\tilde{I}_4=\int \frac{d^nk~d^nl~d^nr~d^ns~~~~r^2(p-l)^2}{k^2l^2(k-l)^2
(r-k)^2(r-l)^2(s-l)^2(r-s)^2(p-r)^2(p-s)^2}\nonumber\\
&&~~~~= \frac{1}{\e}~(p^2)^{1-4\e}~\left[\frac 52\z(5)-\frac 34\z(3)+
{\cal O}(\e)\right]
\label{b4}
\eea
(the explicit evaluation of the last integral was reported in Ref. \cite{us}).

\vspace{0.5cm}
From the previous integrals we can derive all the results needed for our
calculation.\\
By repeated use of (\ref{b1}) we obtain for the $g^0$--order diagram
\beq
\int \frac{d^n q_1 \cdots d^n q_{k-1}}{q_1^2 (q_2-q_1)^2 (q_3-q_2)^2
\cdots (p-q_{k-1})^2} = \frac{1}{\e} \frac{(-1)^k}{[(k-1)!]^2}
(p^2)^{k-2-(k-1)\e} +{\cal O}(1)
\label{b5}
\eeq
For the $g^2$--order (finite) diagram we have
\bea
&&\int d^n q_2 \cdots d^n q_{k-1}
\frac{-q_2^2}{(q_3-q_2)^2 \cdots (p-q_{k-1})^2}
\int \frac{d^nk~d^nl}{k^2 l^2 (k-l)^2 (q_2-k)^2 (q_2-l)^2}\nonumber\\
&&~~~~~~~~~~~= \frac{(-1)^{k-1}(k-1)}{[(k-1)!]^2 k} 12\z(3)(p^2)^{k-2-k\e}
+{\cal O}(\e)
\label{b6}
\eea
In order to obtain this result, one first performs the $k$ and $l$
integrations with the
help of eq. (\ref{b2}), then the other integrals are computed with
the use of eq. (\ref{b1}).

\vspace{0.5cm}
At $g^4$--order, the integrals emerging after having performed the
$D$-algebra are the following ones:\\
For the graph $3a$, by using eq. (\ref{b1}) one obtains
\beq
\int \frac{d^n q_1 \cdots d^n q_{k-1}}{(q_1^2)^{1+2\e} (q_2-q_1)^2
(q_3-q_2)^2 \cdots (p-q_{k-1})^2}
= \frac{1}{\e} \, \frac{(-1)^k(k-1)}{[(k-1)!]^2 (k+1)} (p^2)^{k-2-(k+1)\e}
+{\cal O}(1)
\label{b7}
\eeq
For the graphs $3b$, $4a$ and $4b$, it is easy to deduce
\bea
&& \int d^nr~d^n q_2 \cdots d^n q_{k-1} \frac{1}{(q_2-r)^2 (q_3-q_2)^2
\cdots (p-q_{k-1})^2}
\int \frac{d^nk~d^nl}{k^2 l^2 (k-l)^2 (r-k)^2 (r-l)^2}\nonumber\\
&&~~~~~~~~= \frac{1}{\e} \, \frac{(-1)^k(k-1)}{[(k-1)!]^2 (k+1)}
6\z(3) (p^2)^{k-2-(k+1)\e} +{\cal O}(1)
\label{b8}
\eea
by first evaluating the two--loop integrals in $k$ and $l$ with the
help of
eq. (\ref{b2}), and then applying eq. (\ref{b1}).
\\
The integral relevant for the graph $3c$ is
\bea
&& \int d^n q_3 \cdots d^n q_{k-1} \frac{1}{(q_4-q_3)^2 \cdots
(p-q_{k-1})^2} \nonumber \\
&&~~~~~~~~~\int \frac{d^nk~d^nl~d^nr~d^ns~~~~~~r^2 (q_3-l)^2}
{k^2 l^2 (k-l)^2 (r-k)^2 (r-l)^2 (s-l)^2 (r-s)^2 (q_3-r)^2 (q_3-s)^2}
\nonumber\\
&&~~~~ = \frac{1}{\e} \, \frac{(-1)^k(k-1)}{[(k-1)!]^2 (k+1)}
[6\z(3)-20\z(5)] (p^2)^{k-2-(k+1)\e}+{\cal O}(1)
\label{b9}
\eea
by first evaluating the four--loop integral with momenta $k,l,r,s$ with use
of eq. (\ref{b4}), and then performing the rest of the integrations using
eq. (\ref{b1}).
\\
Finally, the momentum integral for the graph $3d$ is given by
\bea
&& \int d^n q_3 \cdots d^n q_{k-1} \frac{-q_3^2}{(q_4-q_3)^2 \cdots
(p-q_{k-1})^2} \nonumber \\
&&~~~~~~~~~\int \frac{d^nk~d^nl~d^nr~d^ns}
{k^2 l^2 (k-l)^2 (r-k)^2 (s-l)^2 (r-s)^2 (q_3-r)^2 (q_3-s)^2}\nonumber\\
&&~~~~~= \frac{1}{\e}\, \frac{(-1)^k(k-1)}{[(k-1)!]^2 (k+1)}
40\z(5)(p^2)^{k-2-(k+1)\e} +{\cal O}(1)
\label{b10}
\eea
Here, one first performs the $k,l,r,s$ integrations with
eq. (\ref{b3}), and then uses eq. (\ref{b1}).

For the graph $3e$ the momentum integral produced after completion of
the D--algebra is of order $\e$. Indeed, it is given by
\bea
&& \int d^nq_4 \cdots d^nq_{k-1} \frac{1}{(q_5-q_4)^2 \cdots
(p-q_{k-1})^2} \int d^nr r^2(q_4-r)^2 \nonumber \\
&& \int \frac{d^nk~d^nl}{k^2 l^2 (k-l)^2 (r-k)^2 (r-l)^2}
\int \frac{d^ns~d^nt}{(s-r)^2 (t-r)^2 (s-t)^2 (q_4-s)^2 (q_4-t)^2}\nonumber\\
&&~~~~ = \e \,\frac{(-1)^k(k-1)}{[(k-1)!]^2 (k+1)} 144 \z(3)^2
(p^2)^{k-2-(k+1)\e}+{\cal O}(\e^2)
\label{b11}
\eea
This result can be obtained by using first eq. (\ref{b2}) for the $k,l$
and $s,t$ two-loop integrals, and then eq. (\ref{b1}) for the remaining
integrations.

\section{Colour structures}

In this Appendix we give our  conventions and a series of useful
identities involving the group generators. Moreover we show that the
colour structure (\ref{4acolour}) of the nonplanar graphs $4a$ and $4b$
is nonvanishing, by evaluating it explicitly in the $k=4$ case.

For a general simple Lie algebra we have
\beq
[T_a, T_b] = i f_{abc} T_c
\label{c1}
\eeq
where $T_a$ are the generators  and
$f_{abc}$ the structure constants. The matrices $T_a$'s are normalized as
\beq
{\rm Tr}(T_a T_b) =~k_2 ~\d_{ab}
\label{norm}
\eeq
We have also
\beq
f_{amn} f_{bmn}=k_1 \d_{ab}
\label{ff}
\eeq
From the Jacobi identity one obtains
\beq
f_{abm} f_{cdm} + f_{cbm} f_{dam} + f_{dbm} f_{acm} = 0
\label{jacobi}.
\eeq
which in turn allows to write
\beq
f_{alm}f_{bmn}f_{cnl}=\frac{1}{2} k_1 f_{abc}
\label{fff}
\eeq

Now we specialize our formulas to the gauge group $SU(N)$: we have
$k_1=2k_2N$ and we choose
a normalization in (\ref{norm}) such that $k_2=1$.
The generators $T_a$, $a=1, \ldots, N^2-1$, in the fundamental
representation of $SU(N)$
are $N\times N$ traceless matrices.
For $SU(N)$ the relations in (\ref{ff}) and (\ref{fff}) can be
written as
\beq
[[T_a, T_m], T_m] ~=~ 2 N \, T_a
\label{c2}
\eeq
\beq
[[[T_a, T_m], T_n], T_m] ~=~ N \, [T_a, T_n].
\label{c3}
\eeq

Now we concentrate on the explicit evaluation of some colour factors in
the case $k=4$. In particular we want to show that the nonplanar colour
structure of graphs $4a$ and $4b$ is nonvanishing, in contradistinction
 to the case $k=3$ (see formula (A.18) in Appendix A of
\cite{us}).
The relation which allows to deal with products of $T_a$'s is the
following
\beq
T^a_{ij} T^a_{kl}= \left( \d_{il}\d_{jk}-
\frac{1}{N}\d_{ij}\d_{kl}\right).
\label{Tcontract}
\eeq
 From (\ref{Tcontract}) one can obtain several useful formulas
\beq
{\rm Tr}(T_a T_b T_c T_d)~{\rm Tr}(T_a T_b T_c T_d) =  \frac{1}{N^2}
(N^2-1)(N^2+3)
\label{tr1}
\eeq
\beq
{\rm Tr}(T_a T_b T_c T_d)~{\rm Tr}(T_a T_b T_d T_c) = - \frac{1}{N^2}
(N^2-1)(N^2-3)
\label{tr2}
\eeq
\beq
{\rm Tr}(T_a T_b T_c T_d)~{\rm Tr}(T_d T_c T_b T_a) = k_2^4 \frac{1}{N^2}
(N^2-1)(N^4-3N^2+3)
\label{tr3}
\eeq
the last one being a planar type contribution.\\
Moreover, by noticing that
\beq
f_{abc} = -i {\rm Tr} \left( [T_a, T_b] T_c \right)
\eeq
one can compute
\beq
{\rm Tr}(T_{c_1} T_{a_1} T_{c_2} T_{a_2}) f_{c_1mb_1} f_{c_2mb_2} =
- (\d_{b_1a_1}\d_{b_2a_2} + \d_{b_2a_1}\d_{b_1a_2})
\label{tr4}
\eeq
\beq
{\rm Tr}(T_{c_1} T_{c_2} T_{a_1} T_{a_2}) f_{c_1mb_1} f_{c_2mb_2} =
~\d_{b_1b_2}\d_{a_1a_2} + ~{\rm Tr}(T_{b_1} T_{b_2} T_{a_1} T_{a_2})
\label{tr5}
\eeq

\vskip 5mm
Consider now the nonplanar colour structure for the graphs $4a$ and $4b$
\beq
{\cal Q}_k \equiv {\rm Tr} (T_{a_1} \cdots T_{a_k}) \, \sum_{\s} \sum_{i\neq j}
{\rm Tr} \left( T_{a_{\s(1)}} \cdots [[T_{a_{\s(i)}}, T_m], T_n] \cdots
[[T_{a_{\s(j)}}, T_n], T_m] \cdots T_{a_{\s(k)}} \right)
\eeq
As already mentioned ${\cal Q}_3 = 0$. In general however it does not vanish.
In the case $k=4$ by exploiting the various symmetries of
this structure, it can be brought to the more manageable expression
\bea
{\cal Q}_4 &=& 32~[2{\rm Tr}(T_{a_1} T_{a_2} T_{b_3} T_{b_4}) +
{\rm Tr}(T_{a_1} T_{b_3} T_{a_2} T_{b_4})] f_{a_1nm}f_{mpc_1}f_{a_2pr}f_{rnc_2}
\nonumber\\
&&\times [{\rm Tr}(T_{c_1} T_{c_2} T_{b_3} T_{b_4}) +
{\rm Tr}(T_{c_1} T_{b_3} T_{c_2} T_{b_4}) +
{\rm Tr}(T_{c_1} T_{c_2} T_{b_4} T_{b_3})]
\eea
By using the Jacobi identity (\ref{jacobi}) for the four $f$ structure
\beq
f_{a_1nm}f_{mpc_1}f_{a_2pr}f_{rnc_2} = -N f_{a_1mc_1}f_{a_2mc_2}
+ f_{a_1pm}f_{a_2pr}f_{c_1nm}f_{c_2nr}
\eeq
and using equations (\ref{tr4}, \ref{tr5})
and (\ref{tr1}, \ref{tr2}),it is straightforward to obtain
\beq
{\cal Q}_4 = 192  (N^2-1)(2N^2-3)
\label{Q4}
\eeq
As claimed the result is nonvanishing.

\vskip 5mm

Now, as a check, we want to verify in the $k=4$ case that the colour
structure of the diagrams $3c$ and $3d$ is given by the sum of a term
proportional
to the tree level structure ${\cal P}_k$
and a term proportional to the nonplanar structure ${\cal Q}_k$.

The tree level colour structure
\beq
{\cal P}_k \equiv {\rm Tr}(T_{a_1} \cdots T_{a_k}) \, \sum_{\s}
{\rm Tr}(T_{a_{\s(1)}} \cdots T_{a_{\s(k)}})
\eeq
for $k=4$ can be reduced to
\beq
{\cal P}_4 = 4~{\rm Tr}(T_a T_b T_c T_d)~[{\rm Tr}(T_d T_c T_b T_a) +
4~{\rm Tr}(T_a T_b T_d T_c) + {\rm Tr}(T_a T_b T_c T_d)]
\eeq
and then, with (\ref{tr1}--\ref{tr3}), to
\beq
{\cal P}_4 = 4 \frac{1}{N^2}(N^2-1)(N^4-6N^2+18).
\label{P4}
\eeq

Now consider the structure from the diagrams $3c$ and $3d$
\bea
&& {\cal R}_k \equiv {\rm Tr}(T_{a_1} \cdots T_{a_k}) \times \nonumber \\
&& \sum_{\s}
\sum_{i \neq j \neq l} {\rm Tr} \left(T_{a_{\s(1)}} \cdots
[T_{a_{\s(i)}} ,T_m] \cdots [[T_{a_{\s(j)}} ,T_m], T_n] \cdots
[T_{a_{\s(l)}} ,T_n] \cdots T_{a_{\s(k)}} \right)
\ena
Setting $k=4$ it can be written as
\bea
{\cal R}_4 &=& 16~[{\rm Tr}(T_{a_1} T_{a_2} T_{a_3} T_{b}) +
{\rm Tr}(T_{a_1} T_{b} T_{a_2} T_{a_3}) +
{\rm Tr}(T_{a_1} T_{a_2} T_{b} T_{a_3}) +
{\rm Tr}(T_{a_1} T_{a_3} T_{a_2} T_{b}) \nonumber\\
&& + {\rm Tr}(T_{a_1} T_{b} T_{a_3} T_{a_2}) +
{\rm Tr}(T_{a_1} T_{a_3} T_{b} T_{a_2})]
f_{a_1mc_1}f_{a_2rn}f_{nmc_2}f_{a_3rc_3}
[{\rm Tr}(T_{c_1} T_{c_2} T_{c_3} T_{b}) \nonumber\\
&& + {\rm Tr}(T_{c_1} T_{b} T_{c_2} T_{c_3})+
{\rm Tr}(T_{c_1} T_{c_2} T_{b} T_{c_3}) +
{\rm Tr}(T_{c_1} T_{c_3} T_{c_2} T_{b}) +
{\rm Tr}(T_{c_1} T_{b} T_{c_3} T_{c_2}) \nonumber\\
&& + {\rm Tr}(T_{c_1} T_{c_3} T_{b} T_{c_2})]
\eea
Making use of the equations (\ref{tr1}--\ref{tr5}), it is a lengthy
but straightforward calculation to show that
\beq
{\cal R}_4 = 32  (N^2-1)(N^4-18N^2+36).
\label{R4}
\eeq
At this point it is immediate, from eqs. (\ref{Q4}), (\ref{P4}) and (\ref{R4}),
to see that
\beq
{\cal R}_4 = 8 N^2 {\cal P}_4 - {\cal Q}_4
\eeq
in accordance with the manipulations leading to eq. (\ref{3ccolour}).

\end{document}